\documentclass[twocolumn,showpacs,prl,superscriptaddress]{revtex4}
\usepackage{amsmath}
\usepackage{mathrsfs}
\usepackage{graphicx}
\begin{document}

\draft

\title{Anomalous Transmission Phase of a Kondo-Correlated Quantum Dot}%
\author{Kicheon Kang}%
\email{kckang@chonnam.ac.kr}
\affiliation{Department of Physics, Chonnam National University,
  Gwangju 500-757, Korea}%
\author{Mahn-Soo Choi}%
\email{choims@korea.ac.kr}%
\affiliation{Department of Physics, Korea University, Seoul 136-701 Korea}%
\author{Seongjae Lee}%
\affiliation{Basic Research Laboratory, Electronics and
  Telecommunications Research Institute, Daejon 305-350, Korea}%

\date{\today}

\begin{abstract}
We study phase evolution of transmission through a quantum dot with
Kondo correlations. By considering a model that includes nonresonant
transmission as well as the Anderson impurity, we explain 
unusually large phase evolution of about $\pi$ in the Kondo valley observed
in recent experiments.  We argue that this anomalous phase evolution is a 
universal property that
can be found in the high-temperature Kondo phase in the presence of the
time-reversal symmetry.
\end{abstract}

%% 73.23.-b % Electronic transport in mesoscopic systems
%% 73.63.Kv % Quantum Dots
\maketitle

%%%%%%
\let\veps=\varepsilon%
\newcommand\varF{\mathscr{F}}%
\newcommand\eff{\mathrm{eff}}%

Recent measurements of transport through quantum dots have identified
the Kondo effect in a very controllable
way~\cite{gordon98,cronen98,schmid98,simmel99}.  In particular, the
scattering phase shift of the Kondo-assisted transmission has been
measured experimentally~\cite{ji00,ji02,wiel00}.  This measurement has
attracted renewed interest in the Kondo effect since the phase shift cannot
be accessed in bulk Kondo systems nor, even in mesoscopic systems, by
means of conductance measurement.
More importantly, the measured phase shift does not agree with the
theoretical predictions.  The Kondo scattering is expected to induce a
phase shift of $\pi/2$~\cite{langreth66}.  Indeed, theoretical study
based on the impurity Anderson model predicts that the phase evolution
of transmission amplitude should have $\pi/2$-plateaus in the Kondo
limit~\cite{gerland00}.  However, Y. Ji {\em et al.}  observed various
anomalous behavior of the phase evolution~\cite{ji00,ji02} which cannot
be explained in terms of the simple Anderson model.  The experimental
results indicate unusually large span of the phase evolution, such as
the plateaus of the phase shift about $\pi$ in the presence of the Kondo
correlation~\cite{ji00}. Our aim here is to provide a theoretical
explanation on such anomalous phase evolution.

We consider a model that incorporates a weak direct nonresonant 
transmission through a quantum dot (QD), 
as well as the Kondo-resonant transmission. The
importance of including more than the simplest resonant transmission has
been demonstrated in the experiment by Schuster {\em et
  al.}~\cite{schuster97}, which shows unexpected phase lapse by $\pi$ in
the Coulomb blockade (CB) limit. Fano-type interference between the resonant
and the nonresonant transmission provides a possible explanation for the
anomalous phase drop accompanied by transmission
zero~\cite{xu98,deo98,cho98}, in the presence of time-reversal symmetry
(TRS)~\cite{hwlee99}. The role of the nonresonant transmission is expected to
be even more important in the experiment of the Kondo
limit~\cite{ji00,ji02} because the QD should be more open 
to the leads in order
to reach the Kondo limit, which has never been noticed before.  It is
also possible that QD has more than one level contributing to the
transport.  In such a case, the off-resonant transmission plays a similar
role as that of the direct transmission.
Therefore, it is plausible to take into account the nonresonant transmission
through a Kondo-correlated QD, as a first correction to the problem.

In this Letter, we show that some anomalous phase evolution observed in
the experiment can be explained by considering a nonresonant transmission
component interfering with the Kondo resonance.

We begin with the model Hamiltonian
\begin{math}
H = H_L + H_R + H_D + H_T .
\end{math}
The Hamiltonians for the left (L) and right (R) leads are given by
\begin{subequations}
\label{eq:hamil}
\begin{equation}
H_\alpha = \sum_{k\sigma}\varepsilon_{\alpha k}
c^\dagger_{\alpha k\sigma} c_{\alpha k\sigma\alpha} \quad
(\alpha = L, R) \,,
\end{equation}
where $c_{\alpha k\sigma}$ $(c_{\alpha k\sigma}^\dagger)$ is a
destruction (creation) operator of an electron with energy
$\varepsilon_k$, momentum $k$, and spin $\sigma$ on the lead $\alpha$.
The interacting QD is described by
\begin{equation}
H_D = \sum_\sigma \varepsilon_d d_\sigma^\dagger d_\sigma
+ U n_\uparrow n_\downarrow  \,,
\end{equation}
where $d_\sigma$ and $d_\sigma^\dag$ are dot electron operators,
$n_\sigma=d_\sigma^\dag d_\sigma$, $\varepsilon_d$ and $U$ stand for the
energy of the localized level and the on-site interaction, respectively.
The tunneling Hamiltonian $H_T$ has the form
\begin{multline}
\label{kondo-phase:1c}
H_T = \sum_{\alpha=L,R}\sum_{k\sigma}
\left(V_\alpha d_\sigma^\dag c_{\alpha k\sigma} + h.c.\right) \\\mbox{}%
+ \sum_{kk',\sigma}\left(W\, c_{L k\sigma}^\dag c_{R k'\sigma} +
  h.c.\right) \,.
\end{multline}
Here the tunneling amplitude $W$ is responsible for the direct transmission
between the two leads, and $V_\alpha$ for the tunneling between the QD and
the lead $\alpha$.
\end{subequations}

Formally, our model (\ref{eq:hamil}) is equivalent to a two-terminal
Aharonov-Bohm (AB) interferometer containing a 
QD~\cite{bulka01,hofstetter01}, where the
reference arm corresponds (formally) to the term in $W$ in
Eq.~(\ref{kondo-phase:1c}).  However, the previous studies in
Refs.~[\onlinecite{bulka01,hofstetter01}] were focused only on
conductance, whereas our purpose in this work is to investigate the
complex transmission amplitude that contains the phase information as
well as the magnitude.  
Even the magnitude of the nonresonant transmission is in a different
renge from the corresponding values for a typical reference arm
in Refs.~[\onlinecite{bulka01,hofstetter01}].
We also emphasize that the term in $W$ in Eq.~(\ref{kondo-phase:1c})
describes the nonresonant direct transmission through the QD and has a
completely different physical origin from the reference arm in an AB
interferometer.

For simplicity, we assume symmetric junctions (i.e., $V_L=V_R=V$) and
identical leads (i.e.,
$\varepsilon_{Lk}=\varepsilon_{Rk}=\varepsilon_k$) with the density of
states $\rho$ at the Fermi energy.  The direct tunneling matrix element
$W$ is in general complex number, $W=|W|e^{i\varphi}$, while the hopping
matrix elements $V_\alpha$ can be kept as positive real numbers without
loss of generality. Then $\varphi$ stands for the phase difference
between the resonant and the nonresonant component.  We assume the TRS so
that the phase $\varphi$ takes either $0$ or $\pi$.
(In fact, the external magnetic flux penetrating the QD is only a very
small fraction of the flux quantum in the experiment of
Ref.~[\onlinecite{ji00,ji02}], and hence the TRS is well preserved.)

Using the relation between the scattering matrix and the local Green's
function~\cite{langreth66}, one can write the transmission
amplitude of the electrons with energy $\varepsilon$ 
from the left to right lead as
\begin{multline}
t_{LR}(\varepsilon) = ie^{i\varphi}|t_b| \\\mbox{} + ie^{i\varphi}
\Gamma_{\eff} G_d^R(\varepsilon) \left[ |r_b|\cos{\varphi} -
  i(|t_b|+\sin{\varphi}) \right] \,.
\label{eq:trans2}
\end{multline}
Here $|t_b|\equiv 2x/(1+x^2)$ with $x=\pi\rho|W|$ is the magnitude of
the direct transmission amplitude.  $|r_b|$ is defined by the relation
$|t_b|^2+|r_b|^2=1$.  The effective hybridization parameter
$\Gamma_\eff$ in Eq.~(\ref{eq:trans2}) is defined by
\begin{math}
\Gamma_\eff = \Gamma/(1+x^2)
\end{math}
with $\Gamma = 2\pi\rho V^2$, and $G_d^R(\varepsilon)$ is the retarded
Green's function for the dot electron.

At zero temperature, only the electrons at the Fermi energy contribute to
the total transmission amplitude ($t_{LR}$), and 
the Friedel-Langreth sum rule~\cite{langreth66}
gives an exact expression for $G_d^R$ in terms of the occupation number
of the dot, $n_d$, leading to the relation
\begin{subequations}
\label{eq:trans-FL}
\begin{gather}
t_{LR}= t_{LR}(0)
= \frac{ie^{i\varphi}|t_b|}{e_d-i} \left( e_d+Q \right) \,,\\
e_d = \cot{(\pi n_d/2)} \,,\\
Q = -\frac{|r_b|}{|t_b|} \cos{\varphi} + \frac{i}{|t_b|} \sin{\varphi}
\,.
\end{gather}
\end{subequations}
Equation~(\ref{eq:trans-FL}) already provides some important
informations. First, transmission zero takes place at $\cot{(\pi n_d/2)}
=\pm|r_b|/|t_b|$ for $\varphi=0,\pi$, as a result of destructive
interference between the two transmission components.  For $|t_b|\ll 1$,
transmission zero is located far from the Kondo limit, $n_d\simeq0$ or
$n_d\simeq2$, for $\varphi=0$ or $\varphi=\pi$, respectively. In the
opposite limit ($|t_b|\simeq 1$), $t_{LR}$ goes to zero in the Kondo
limit ($n_d\simeq 1$). This limit was investigated previously for a
ballistic quantum wire coupled to a QD~\cite{kang01}.

At finite temperatures, we need to take the thermal average of the
transmission amplitude~\cite{endnote:1}:
\begin{equation}
\label{eq:trans-finiteT}
t_{LR} =
\int \left( -\frac{\partial f}{\partial\varepsilon} \right)
t_{LR}(\varepsilon)\, d\varepsilon \;,
\end{equation}
where $f$ denotes the Fermi distribution function.
%% This expression can be achieved by extracting the interference term
%% of the thermally averaged transmission probability in a `two-slit' type
%% interferometer. 
%%

For a quantitative study, we will adopt the slave-boson mean field
theory (SBMFT) assuming $U=\infty$~\cite{hewson93}.  We will also do the
numerical renormalization group (NRG) calculations to confirm the
results from the SBMFT.
The SBMFT satisfies the unitarity of the scattering
matrix~\cite{kang01}, which cannot be preserved in some other approaches
based on the $1/N_s$-expansion (with $N_s$ being the degeneracy of the
level). After some algebra, we obtain the relation
\begin{subequations}
\label{kondo-phase::eq:t:SBMFT}
\begin{gather}
t_{LR}(\varepsilon) = \frac{ie^{i\varphi}|t_b|}{\tilde{e}_d-i} \left(
  \tilde{e}_d+Q \right), \\
\tilde{e}_d \equiv \frac{\tilde{\varepsilon}_d-\varepsilon}{(1-n_d)
  \Gamma_{\eff}} \,.
\end{gather}
\end{subequations}
The renormalized energy level $\tilde\varepsilon_d$ in
Eq.~\eqref{kondo-phase::eq:t:SBMFT} will be determined self-consistently
together with $n_d$.  We note that at $T=0$, the
expression in Eq.~(\ref{kondo-phase::eq:t:SBMFT}) based on the SBMFT
reduces to the exact form of Eq.~(\ref{eq:trans-FL}).

\begin{figure}
\begin{center}
\includegraphics*[width=40mm]{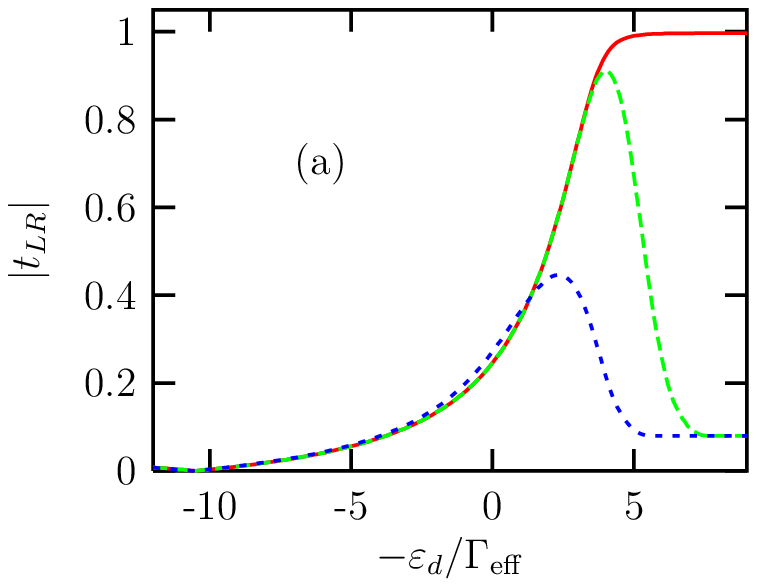}
\includegraphics*[width=40mm]{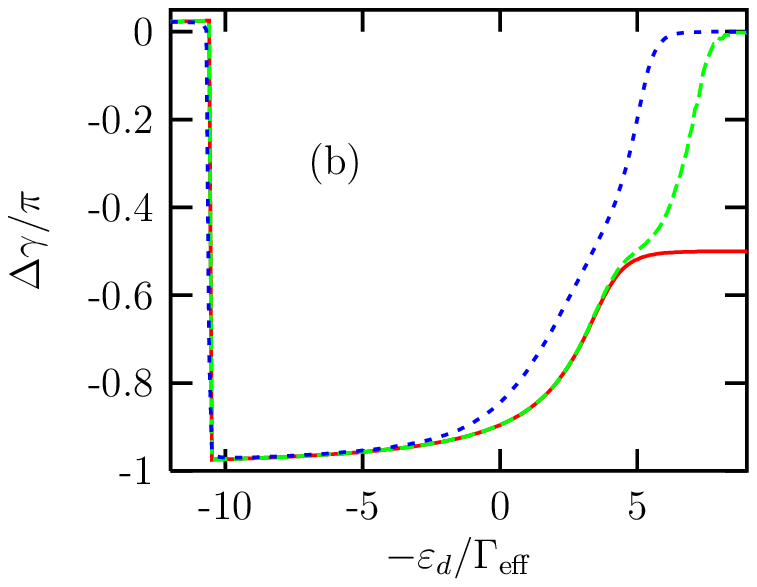}\\%
\includegraphics*[width=40mm]{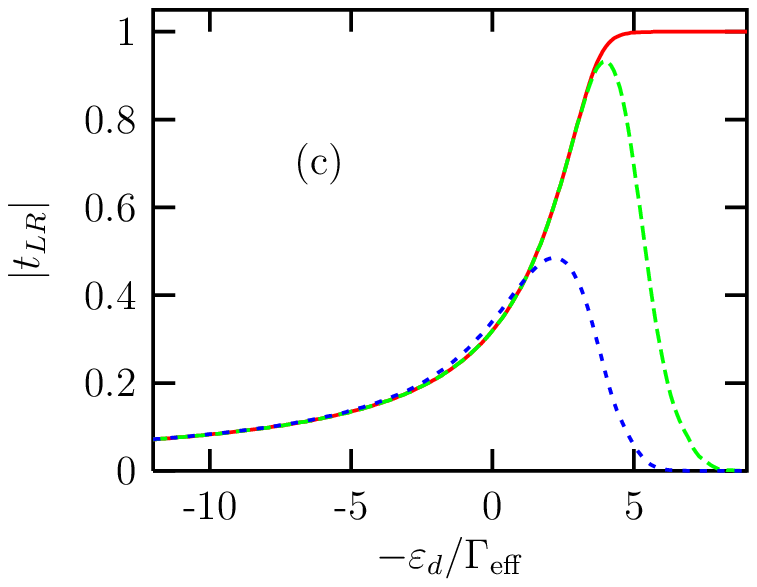}
\includegraphics*[width=40mm]{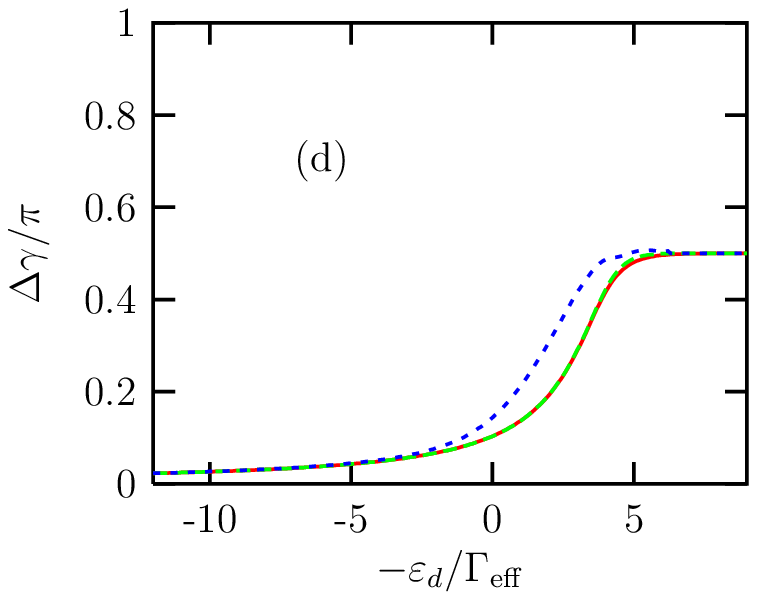}%
 \caption{(a) The magnitude and (b) the phase of the transmission amplitude
   $t_{LR}$ for $\varphi=0$ and $|t_b|=0.08$ with the temperatures 
   $T=0$ (solid lines),
   $0.02\Gamma_{\eff}$ (dashed lines), $0.5\Gamma_{\eff}$ (dotted
   lines).  (c) The magnitude and (d) the phase for $t_b=0$ with the
   temperatures $T=0$ (solid lines), $0.02\Gamma_{\eff}$ (dashed
   lines), $0.5\Gamma_{\eff}$ (dotted lines).}
\label{fig1}
\end{center}
\end{figure}

The results from the SBMFT are summarized in Fig.~\ref{fig1} for 
$\varphi=0$~\cite{endnote:2}.
Figure~\ref{fig1} shows (a) the magnitude $|t_{LR}|$ and (b) the phase shift
$\Delta\gamma$ of the total transmission amplitude $t_{LR}$ at several
temperatures in the presence of a small direct transmission
($|t_b|=0.08$).  For a comparison, the results for $t_b=0$ 
 are also shown in Fig.~\ref{fig1} (c) and (d).  One can
see clearly that while the magnitude is affected very little, a small
$|t_b|$ can lead to completely different behavior of the phase at finite
temperatures, as we discuss in detail now.

According to the behaviors of the transmission phase in the presence of
direct transmission, the low temperature region can be divided into two
sub-regions: the ``unitary Kondo regime'' ($T<T_0$), and the so-called
``Fano-Kondo regime'' ($T>T_0$)~\cite{temp-higher}: 
see below for an estimate of the
crossover temperature $T_0$.
In the unitary Kondo regime, the Kondo resonance provides a
transmission channel with a transmission probability larger than
the direct transmission $|t_b|^2$.  Therefore, neither the magnitude nor
the phase of $t_{LR}$ is affected by the small $|t_b|$.  Namely, as well
understood by the studies based on the Anderson impurity
model~\cite{glazman88,ng88}, $|t_{LR}|$ ($\Delta\gamma$) changes from 0
to 1 ($\pi/2$) as $\varepsilon_d$ varies from the empty dot limit
($\varepsilon_d\gg\Gamma_\eff$) to the singly occupied limit
($\varepsilon_d\ll-\Gamma_\eff$).

In the Fano-Kondo regime, on the other hand, one can observe
much richer behaviors.  As the temperature increases, the Kondo effect
is partially suppressed and the transmission probability through the Kondo
resonance becomes comparable to the nonresonant transmission $|t_b|^2$.
There occurs an interference between the nonresonant transmission and the
transmission through the Kondo resonance.  Such a Fano-type interference
affects $|t_{LR}|$ very little, since the nonresonant transmission and the
transmission through the Kondo resonance are both already small in the
region where the interference is important; compare Figs.~\ref{fig1} (a)
and (c).
However, the phase shift ($\Delta\gamma$) is affected
significantly by even a small value of $|t_b|$.  As shown in
Fig.~\ref{fig1} (b), the plateau of $\Delta\gamma$ as a function of
$\varepsilon_d$ is lifted significantly from $\pi/2$ to a value close to
$\pi$.  This behavior is consistent with the experimental
observation~\cite{ji00}, but is in a strong contrast with the almost
temperature independent Kondo plateaus at $\pi/2$ for $t_b=0$
[Fig.~\ref{fig1}(d)].  We believe that this anomalous phase behavior in
the presence of the nonresonant transmission can be a natural explanation
observed in the experiment~\cite{ji00}, that is, the phase evolution of
about $\pi$ in the Kondo valley~\cite{silvestrov03}.

\begin{figure}
\begin{center}
\includegraphics*[width=60mm]{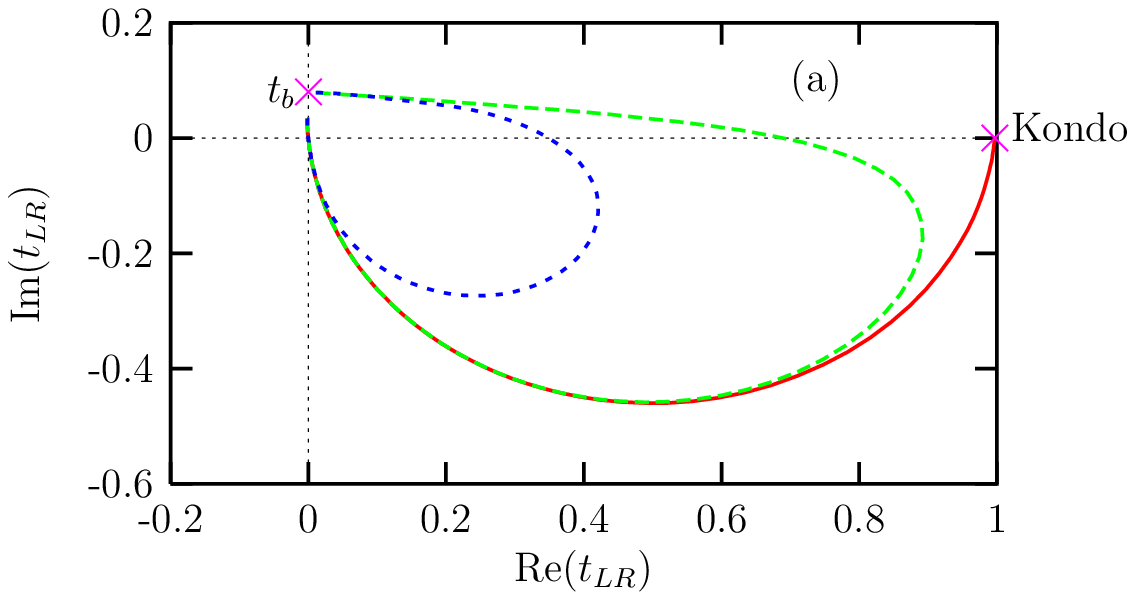}\\%
\includegraphics*[width=60mm]{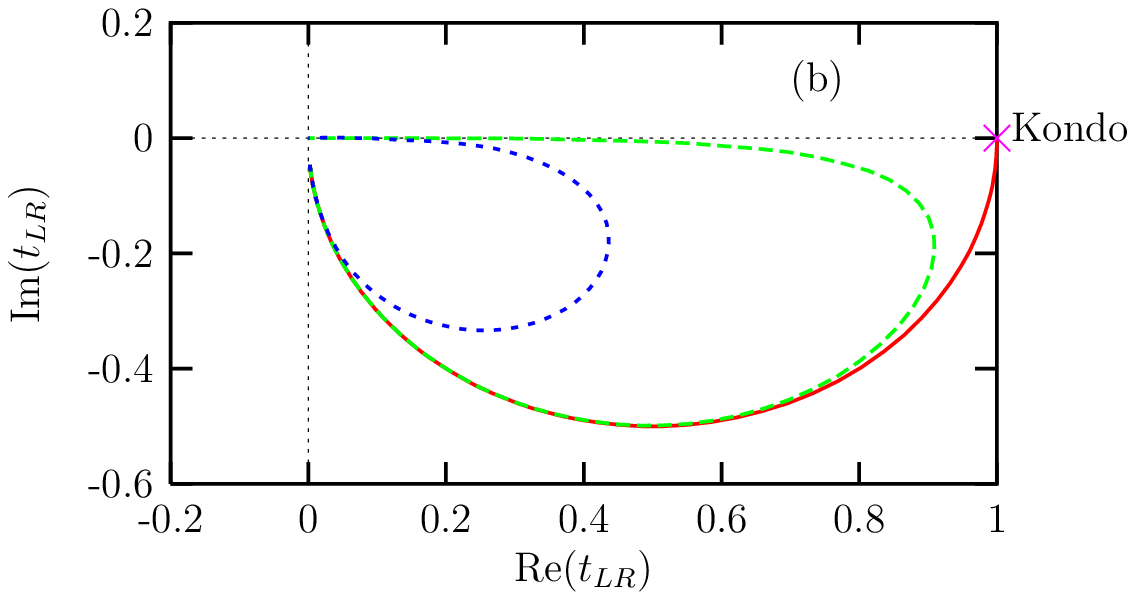}%
\caption{Trajectories in the complex plane for
  the numerically obtained transmission amplitudes in Fig.~\ref{fig1}
  for (a) $|t_b|=0.08$, $\varphi=0$, and for (b) $t_b=0$.}
\label{fig2}
\end{center}
\end{figure}

In fact, this unexpected behavior of transmission phase is better understood
by investigating the trajectories of the
transmission amplitude $t_{LR}$ in the complex plane as $\varepsilon_d$
varies from $\varepsilon_d\gg\Gamma_\mathrm{eff}$ to 
$\varepsilon_d\ll -\Gamma_\mathrm{eff}$ 
at different temperatures for $t_b\neq 0$ (and also for $t_b=0$); 
see Fig.~\ref{fig2}. Notice that the following argument is quite universal
that relies only on the existence of nonresonant transmission and the TRS.
The most important change due to the the direct transmission is that the
transmission coefficient has a finite value, $t_{LR} = t_b$, even when
the resonant transmission component is suppressed.  This put
a negligible effect on $t_{LR}$ at $T<T_0$, where the resonant
transmission component is not suppressed and larger in magnitude
than the direct transmission component.  But it plays a significant role
in the Fano-Kondo regime, where the Kondo-assisted transport is
partially suppressed.  The suppression of the Kondo-assisted
transmission leads to $\Delta\gamma$ significantly larger than $\pi/2$,
even close to $\pi$, since $t_b$ has pure imaginary value in the
presence of TRS (i.e., $t_b=ie^{i\varphi}|t_b|$ and
$\varphi=0,\pi$).

\begin{figure}
\centering%
\includegraphics*[width=60mm]{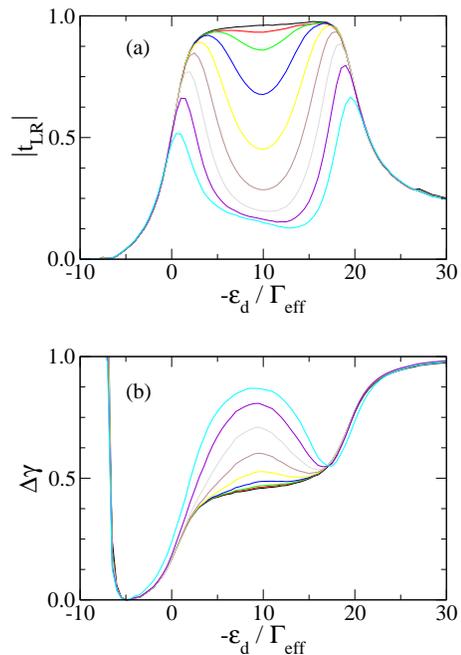}
\caption{The same as Fig.~\protect\ref{fig1} (a) and (b),
  except that the results are from the NRG calculations and for finite
  $U$.  $U=20\Gamma_\eff$, $W=0.065D$ ($|t_b|=0.2$), $V=0.2D$, and
  $T/\Gamma_\eff=0, 10^{-4}, 10^{-3.5}, 10^{-3}, \cdots, 10^{-0.5}$ from
  top to bottom in (a) and bottom to top in (b).}  
\label{kondo-phase::fig3}
\end{figure}

So far we have discussed the results based on the SBMFT for $U=\infty$.
We stress that our findings about the unusual phase evolution are quite
universal, which do not depend on the approximations adopted here nor on
the constraint of $U=\infty$.  To confirm this, we also provide the
results from the NRG calculations in Fig.~\ref{kondo-phase::fig3}, which
are in good agreement with those from the SBMFT except that now there
is a region where the dot is doubly occupied ($\varepsilon_d<-U$).  
Further, the results show clearly the crossover from the unitary to the
Fano-Kondo region as temperature increases.

\paragraph{Estimation of the crossover temperature $T_0$:}%
We now estimate the crossover temperature $T_0$.  $T_0$ can be
determined by comparing $|t_b|$ and the magnitude of the resonant
component. That is, crossover from the unitary to the Fano-Kondo phase
takes place at the temperature where the magnitude of the resonant
transmission is comparable to $|t_b|$.
Since the Kondo-correlated state behaves like a Fermi liquid, we
substitute
\begin{equation}
\label{kondo-phase::eq:Gd}
G_d^R(\varepsilon) \approx
\frac{T_K/\Gamma_{\eff}}{\varepsilon+iT_K} 
\end{equation}
into Eqs.~(\ref{eq:trans2}) and (\ref{eq:trans-finiteT}), and find that
(for $\varphi=0$)
\begin{equation}
\label{eq:tlr-X}
t_{LR} = i|t_b| + i\left( |r_b| - i|t_b| \right)
\varF(T_K) \,,
\end{equation}
where
\begin{equation}
\varF(T_K) =  \int{d\veps}\;
  \left( -\frac{\partial f}{\partial\varepsilon} \right)
  \frac{T_K}{\varepsilon+iT_K} \;. \label{eq:X}
\end{equation}
The integral in Eq.(\ref{eq:X}) can be calculated exactly with the help
of contour integration~\cite{bloomfield67}, which leads in the limit of
$T\gg T_K$ to the form
\begin{equation}
\varF(T_K) \simeq -i\frac{T_K}{T_K+\pi T} \;.
\end{equation}
Inserting this expression into Eq.(\ref{eq:tlr-X}), one can find that
the crossover from $\Delta\gamma=\pi/2$ to $\Delta\gamma=\pi$ takes
place at $T\sim T_0$ such that
\begin{equation}
\label{eq:T1}
T_0 = \frac{|r_b|}{\pi|t_b|} \min(T_K) \,,
\end{equation}
where $\min(T_K) = T_K(\varepsilon_d=-U/2)$.

Equation~(\ref{eq:T1}) is useful to test our claims.
We recall that $T_K$ can be extracted from the temperature dependence of
the conductance, and $|t_b|$ from the Fano-resonance shape of the
conductance at higher temperatures.  Equation~\eqref{eq:T1} then
estimates $T_0$.  One has only to compare $\Delta\gamma$ as a function
of $\veps_d$ at $T\ll T_0$ and $T>T_0$.
We add that $T_0$ is slightly overestimated in Eq.~(\ref{eq:T1}) since
the Fermi liquid form has been used for estimation even at finite
temperatures.
The $\pi$-plateaus in the Fano-Kondo regime were observed
experimentally~\cite{ji00,ji02}, according to our interpretation.
In the same experiments, however, the $\pi/2$-plateaus in the unitary
Kondo limit were not obverved.
We point out that in those experiments, the dot was too open and in the
mixed valence regime (instead of the unitary Kondo limit) for
strong coupling between the leads and the QD.
%% In the experiment of Ref.~\onlinecite{ji00,ji02}
%% shows only the $\pi$-plateaus in the Fano-Kondo regime. In our opinion,
%% the reason why the $\pi/2$-plateaus was not observed is because the charge
%% fluctuation was significant in the unitary limit for this experiment.

In conclusion, we have theoretically explained unusually large value of
the transmission phase ($\sim\pi$) found in a recent experiment for the
Kondo regime of a quantum dot. For the Anderson impurity as well as the
nonresonant transmission between the two leads, we found that
time-reversal symmetry at high-temperature Kondo phase results in the
plateaus of about $\pi$, as long as the nonresonant transmission is small
but finite.

\acknowledgements%
We acknowledge helpful discussions with 
Y.~Ji, H.-W.~Lee, and P.~Silvestrov.  This work 
was supported by the KMIC, the SKORE-A,
the eSSC at Postech, and a Chonnam National University Grant.  M.-S.C.\ 
acknowledges the support from a KRF Grant (KRF-2002-070-C00029).

%%%%%% References


\begin{thebibliography}{99}
\bibitem{gordon98} D. Goldhaber-Gordon, H. Shtrikman, D. Abush-Magder,
  U. Meirav and M. A. Kastner, Nature {\bf 391}, 156 (1998); D.
  Goldhaber-Gordon {\em et al.}, Phys. Rev. Lett. {\bf 81}, 5225 (1998).
 
\bibitem{cronen98} S. M. Cronenwett, T. H. Oosterkamp, and L. P.
  Kouwenhoven, Science {\bf 281}, 540 (1998).
  
\bibitem{schmid98} J. Schmid {\em et al.}, Physica {\bf 256B-258B}, 182
  (1998).
  
\bibitem{simmel99} F. Simmel {\em et al.}, Phys. Rev. Lett. {\bf 83},
  804 (1999).
 
\bibitem{ji00} Y. Ji, M. Heiblum, D. Sprinzak, D. Mahalu, and H.
  Shtrikman, Science {\bf 290}, 779 (2000).
 
\bibitem{ji02} Y. Ji, M. Heiblum, and H. Shtrikman, Phys. Rev. Lett.
  {\bf 88}, 076601 (2002).
 
\bibitem{wiel00} W. G. van der Wiel, S. De Franceschi, T. Fujisawa, J.
  M.  Elzerman, S. Tarucha, and L. P. Kouwenhoven, Science {\bf 289},
  2105 (2000).
  
\bibitem{langreth66} D. C. Langreth, Phys. Rev. {\bf 150}, 516 (1966).
  
\bibitem{gerland00} U. Gerland, J. von Delft, T. A. Costi, and Y. Oreg,
  Phys. Rev. Lett. {\bf 84}, 3710 (2000).
  
\bibitem{schuster97} R. Schuster, E. Buks, M. Heiblum, D. Mahadu, V.
  Umansky and H. Shtrikman, Nature {\bf 385}, 417 (1997).
  
\bibitem{xu98} H. Xu and W. Sheng, Phys. Rev. B {\bf 57}, 11903 (1998).
  
\bibitem{deo98} P. S. Deo, Solid State Comm. {\bf 107}, 69 (1998).
  
\bibitem{cho98} C.-M. Ryu and S. Y. Cho, Phys. Rev. B {\bf 58}, 3572
  (1998).
  
\bibitem{hwlee99} H.-W. Lee, Phys. Rev. Lett. {\bf 82}, 2358 (1999).
  
\bibitem{bulka01} B. R. Bulka and P. Stefa\'nski, Phys. Rev. Lett. {\bf
    86}, 5128 (2001).
  
\bibitem{hofstetter01} W. Hofstetter, J. K\"onig, and H. Schoeller,
  Phys. Rev. Lett. {\bf 87}, 156803 (2001).
  
\bibitem{kang01} K. Kang, S. Y. Cho, J.-J. Kim, and S.-C. Shin, \prb
  {\bf 63}, 113304 (2001).
 
\bibitem{endnote:1} We have in mind an experiment on AB
  interferometer with our system put in one of the two arms. 
  Eq.~(\ref{eq:trans-finiteT}) actually accounts only for the interference
  term when multiple backscatterings are negligible (see 
  Ref.~\onlinecite{schuster97})
  
\bibitem{hewson93} For a review, see. e.g., A. C. Hewson, {\em The Kondo
    Problem to Heavy Fermions} (Cambridge University Press, Cambridge
  1993).
 
\bibitem{endnote:2}%
  In this work we mainly focus on the case of $\varphi=0$.  One can show
  that the Hamiltonian is invariant under the electron-hole
  transformation together with the change $\varphi=\pi \to 0$; i.e., the
  results for $\varphi=\pi$ can be deduced from those for $\varphi=0$.

\bibitem{temp-higher}%
  At even higher temperatures (the CB region), 
  \emph{phase lapses} by $\pi$ were observed experimentally.  
  While this effect is
  partially understood for non-interacting models, there is no
  theoretical explanation for strongly correlated systems in question.
  %%
  In this work we are only confined in the region where the temperature
  is much lower than such a crossover region.

\bibitem{glazman88} L. I. Glazman and M. E. Raikh, Pis'ma Zh. Eksp.
  Teor. Fiz. {\bf 47}, 378 (1988) [JETP Lett. {\bf 47}, 452 (1988)].
 
\bibitem{ng88} T. K. Ng and P. A. Lee, Phys. Rev. Lett. {\bf 61}, 1768
  (1988).
  
\bibitem{silvestrov03} Actually, there is a different interpretation
  based on scaling argument about the plateau in the Kondo valley at
  high temperatures.  See, P. G. Silvestrov and Y. Imry,
  Phys.~Rev.~Lett.  {\bf 90}, 106602 (2003).  However, they did not
  explain why the phase shift is not $\pi/2$.
  
\bibitem{bloomfield67} P. E. Bloomfield and D. R. Hamann, Phys. Rev.
  {\bf 164}, 856 (1967).

\end{thebibliography}
\end{document}